\documentclass[aps,prl,reprint,twocolumn,showpacs,superscriptaddress]{revtex4-2}

\usepackage{amsmath}
\usepackage{graphicx}
\usepackage{lmodern}
\usepackage{amsmath}
\usepackage{color}
\usepackage{amssymb}
\usepackage{bm}
\usepackage{braket}
\usepackage{mathtools}

\newcommand{\br}{\bm{r}}

\newcommand{\bq}{\bm{q}}
\newcommand{\bQ}{\bm{Q}}

\usepackage[normalem]{ulem} 
\newcommand{\ymw}[1]{{\color[rgb]{0, 0, 0}{#1}}}

\usepackage[dvipsnames]{xcolor}
\usepackage[colorlinks=true]{hyperref}
\hypersetup{
    colorlinks=true,
    linkcolor=blue,
    citecolor=blue,
    urlcolor=blue
}

\makeatletter
\def\maketitle{
\@author@finish
\title@column\titleblock@produce
\suppressfloats[t]}
\makeatother

\begin{document}
\title{Origin of Spin Stripes  in Bilayer Nickelate La$_3$Ni$_2$O$_7$}

\author{Hao-Xin Wang}
\affiliation{Department of Physics, The Chinese University of Hong Kong, Sha Tin, New Territories, Hong Kong, China}

\author{Hanbit Oh}
\affiliation{William H. Miller III Department of Physics and Astronomy, Johns Hopkins University, Baltimore, Maryland, 21218, USA}

\author{Tobias Helbig}
\affiliation{Leinweber Institute for Theoretical Physics, Stanford University, Stanford, CA 94305, USA}

\author{Bai Yang Wang}
\affiliation{Stanford Institute for Materials and Energy Sciences, SLAC National Accelerator Laboratory,
Menlo Park, CA 94025, USA}
\affiliation{Departments of Applied Physics, Stanford University, Stanford, CA 94305, USA}

\author{Jiarui Li}
\affiliation{Stanford Institute for Materials and Energy Sciences, SLAC National Accelerator Laboratory,
Menlo Park, CA 94025, USA}
\affiliation{Departments of Applied Physics, Stanford University, Stanford, CA 94305, USA}

\author{Yijun Yu}
\affiliation{Stanford Institute for Materials and Energy Sciences, SLAC National Accelerator Laboratory,
Menlo Park, CA 94025, USA}
\affiliation{Departments of Applied Physics, Stanford University, Stanford, CA 94305, USA}

 \author{Harold Y. Hwang}
\affiliation{Stanford Institute for Materials and Energy Sciences, SLAC National Accelerator Laboratory,
Menlo Park, CA 94025, USA}
\affiliation{Departments of Applied Physics, Stanford University, Stanford, CA 94305, USA}

\author{Hong-Chen Jiang}
\affiliation{Stanford Institute for Materials and Energy Sciences,
SLAC National Accelerator Laboratory, Menlo Park, CA 94025, USA}

\author{Yi-Ming Wu}
\email{yimwu@stanford.edu}
\affiliation{Leinweber Institute for Theoretical Physics, Stanford University, Stanford, CA 94305, USA}
\affiliation{Stanford Institute for Materials and Energy Sciences, SLAC National Accelerator Laboratory,
Menlo Park, CA 94025, USA}

\author{S. Raghu}
\email{sraghu@stanford.edu}
\affiliation{Leinweber Institute for Theoretical Physics, Stanford University, Stanford, CA 94305, USA}
\affiliation{Stanford Institute for Materials and Energy Sciences, SLAC National Accelerator Laboratory,
Menlo Park, CA 94025, USA}

\begin{abstract}
\ymw{The bilayer nickelate La$_3$Ni$_2$O$_7$ has recently emerged as a high-temperature superconductor with unusual spin stripe order in its ambient pressure phase. We propose a microscopic Hamiltonian that faithfully reflects the crystalline symmetry of this system, with the primary aim of addressing its unconventional magnetism at ambient pressure. Using state-of-the-art density matrix renormalization group calculations, we show that $(\pi/2,\pi/2)$ spin stripe order arises in our model at sizable Hund's coupling $J_H$ from a hidden quasi-one dimensionality and persists over a range of electron concentrations. In the more symmetric high-pressure regime,  our model exhibits enhanced interlayer pairing tendencies when the interlayer antiferromagnetic coupling $J_\bot$ becomes sufficiently large. Our results provide a microscopic origin of the diagonal spin stripes and identify Hund’s coupling $J_H$ and interlayer  coupling $J_\bot$ as key ingredients governing magnetic order and pairing tendencies in La$_3$Ni$_2$O$_7$.
}

 \end{abstract}

\maketitle

\date{\today}

{\it Introduction.}--The Ruddlesden–Popper bilayer nickelate La$_3$Ni$_2$O$_7$ has emerged as a new type of high-$T_c$ superconductor under high hydrostatic pressure, exhibiting transition temperatures as high as $80$K~\cite{Sun2023,Zhang2024exp,Hou_2023,Liu2024,Yang2024,ZHANG2024147}. At ambient pressure, an unusual type of spin stipe order (SSO) with momentum $\bQ=(\pi/2,\pi/2)$ and onset temperature $T_\text{SSO}\approx150$K is observed instead, through various measurements including resonant inelastic X-ray scattering~\cite{Chen2024}, $\mu$SR~\cite{PhysRevLett.132.256503,Khasanov2025}, NMR~\cite{ZHAO20251239,kakoi2024multiband}, and inelastic neutron scattering~\cite{XIE20243221}. 
More recently, strain engineered thin films of La$_3$Ni$_2$O$_7$ at ambient pressure have also been shown to be superconducting ($T_c\approx 40$K) only if a sufficiently compressive strain is exerted~\cite{Ko2024,Zhou2025,Liu2025}, while without strain, the thin film sample again exhibits the  $\bQ=(\pi/2,\pi/2)$ SSO with similar $T_\text{SSO}$ as in the bulk case~\cite{Gupta2025,Ren2025}. The thin film samples allow for easier access to the superconducting properties of this material, including the electronic structure from angle resolved photoemission spectroscopy(ARPES)~\cite{Li_2025,yue2025correlatedelectronicstructuresunconventional,Shao2025BandThinFilm,wang2025electronicstructurecompressivelystrained} and scanning tunneling microscopy (STM)~\cite{fan2025superconductinggapstructurebosonic}.
However, the pairing symmetry remains under debate, with competing proposals of $s_\pm$-wave and $d$-wave states\cite{fan2025superconductinggapstructurebosonic,guo2025revealingsuperconductinggap,cao2025directobservationdwave,Shao2025PairingWithoutGamma}.

\begin{figure}
    \includegraphics[width=8.5cm]{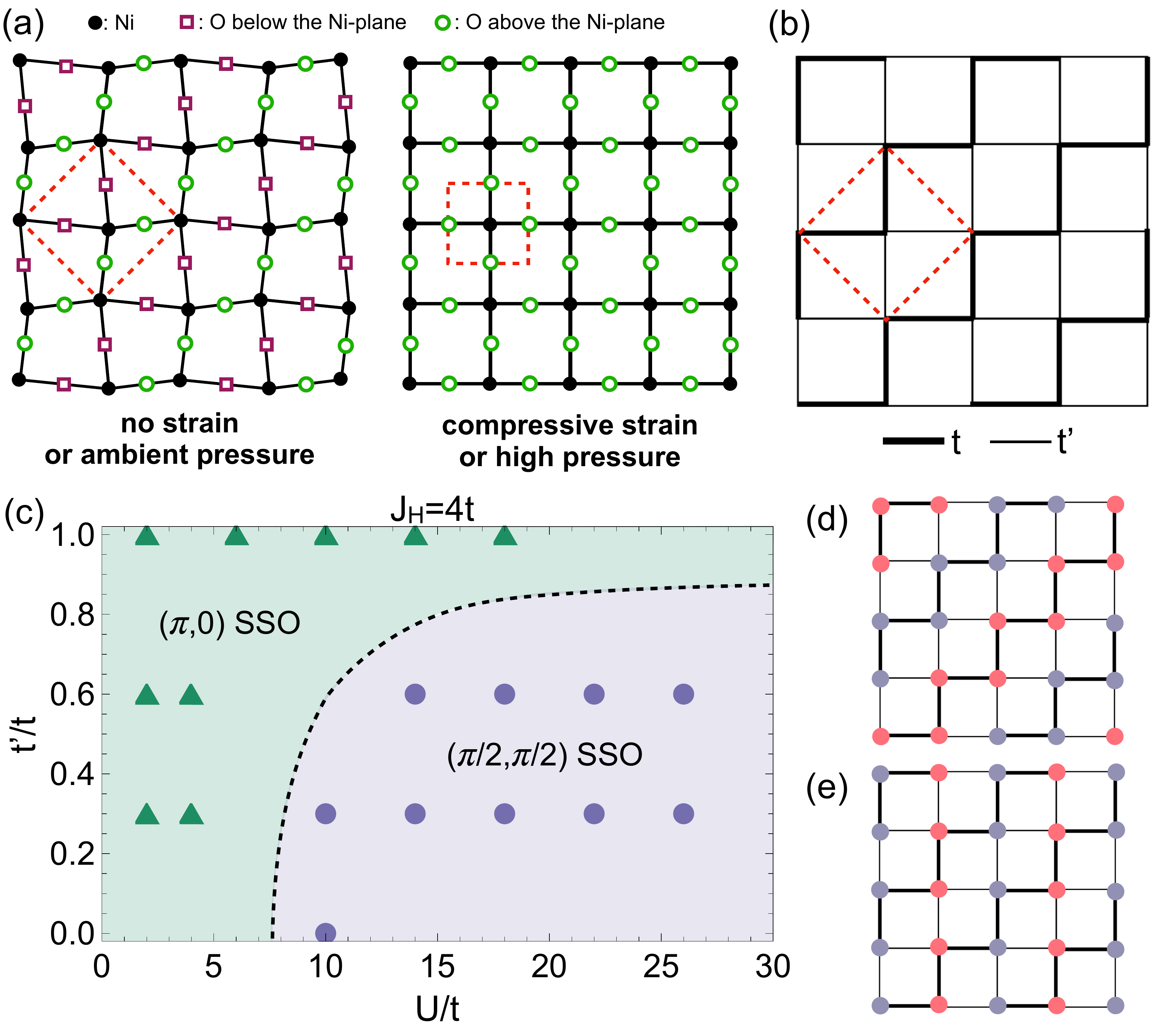}
    \caption{(a) Illustration of a single Ni-O layer. At ambient pressure, the  Ni sites are distorted from a square lattice, while the O atoms are buckled either above or below the Ni plane. At high pressure, the Ni sites form a square lattice, with all O's buckled in the same direction. (b) In a minimal model that includes only the two $e_g$ orbitals from Ni, the pressure/strain effect can be modeled by adding some anisotropy into the $d_{x^2-y^2}$ hopping, resulting in both strong ($t$) and weak ($t'$) bonds. (c) Phase diagram of the model in Eq.\eqref{eq:model} as a function of $t'$ and $U$, for a particular $J_H=4t$ and $J_\bot=0$ obtained from DMRG calculation.  The $(\pi/2,\pi/2)$ and $(\pi,0)$ SSOs are shown in (d) and (e) respectively. Adding a small $J_\bot$ merely flips the spin patterns  between the two layers.  }
    \label{fig:model}
\end{figure}

Two indispensable ingredients in La$_3$Ni$_2$O$_7$ are  essential for understanding the experimentally observed physics. The first one is the multi-orbital character due to the $3d^{7.5}$ electron configuration of the Ni$^{2.5+}$ cations: the low energy electrons are from nickel $d_{z^2}$ and $d_{x^2-y^2}$ orbitals, and the Hund's coupling $J_H$ between them introduces another energy scale in addition to the onsite Coulomb repulsion. 
The second ingredient is the pressure/strain induced structural transition. 
At ambient pressure (or without strain for the thin film),  La$_3$Ni$_2$O$_7$ has the space group $Amam$. A slice of the Ni-plane in this case is illustrated in Fig.~\ref{fig:model}(a), where the oxygen atoms are buckled  above and below the Ni-plane and the Ni-Ni bond length is non-uniform, {resulting in a doubled unit cell.}
At higher pressure, the bulk structure changes to $Fmmm$, which is orthorombic but nearly tetragonal, and may further change to tetragonal $I4/mmm$ at even higher pressure~\cite{Wang2024,GeislerStruct2024}, while for the thin film sample with compressive strain $I4/mmm$ symmetry is enforced by substrate.
Ref.~\cite{bhatt2025resolvingstructuraloriginssuperconductivity} shows that both the bulk sample at high pressure and the thin film sample with compressive strain  can be illustrated as in the right panel of Fig.~\ref{fig:model}(a), where a single Ni layer forms a regular square lattice, and all the nearby oxygens are buckled in a uniform manner. 
Given that superconductivity is observed only in the more symmetric {high pressure (compressively strained) region}, while the spin stripe develops in the less symmetric {ambient pressure regime}, it is crucial to understand the nature of these two regimes
in a  unified manner.

Although there have been proposals on the mechanism of the superconducting phase~\cite{PhysRevB.108.L140505,PhysRevB.108.L201108,PhysRevLett.132.146002,PhysRevLett.132.036502,PhysRevB.108.174511,PhysRevB.110.104517,PhysRevLett.131.126001,PhysRevB.111.174506,shen2023effective,Wu2024,PhysRevB.111.L020504,PhysRevLett.131.206501,PhysRevLett.131.236002,PhysRevB.109.L081105,lu2023superconductivity,PhysRevLett.132.126503,PhysRevB.109.165154,PhysRevLett.133.126501,PhysRevB.108.L140504,PhysRevB.108.174501,PhysRevB.108.165141,2023arXiv230812750K,fabian1,fabian2,fabian3,zhan2024cooperation,PhysRevMaterials.8.074802,wang2024pressure,Xue_2024,Zhang2024,jiang2024high,PhysRevB.108.214522,oh2025highspinlowspin,Geisler2024,geisler2025,khaliullin2025,oh2025dopingspinonemottinsulator,PhysRevB.111.L241102,PhysRevB.108.L180510,PhysRevB.110.094509,zhu2025quantumphasetransitiondriven,PhysRevB.110.L060510,Ji2025StrongCoupling,Shao2026ElectricField,Jialin2024Orbital}, a theoretical understanding of  the $\bQ=(\pi/2,\pi/2)$ spin order has remained a challenge~\cite{PhysRevLett.108.126406,liu2025origindiagonaldoublestripespindensitywave,wang2024electronicmagneticstructuresbilayer}.
\ymw{In this paper, we focus on elucidating the microscopic origin of the spin stripe order in La$_3$Ni$_2$O$_7$ at ambient pressure.}
We model the $Amam$ phase by a hopping non-uniformity $t'\neq t$ for the $d_{x^2-y^2}$ orbitals,
as shown in Fig.~\ref{fig:model}(b), and approximate the $d_{z^2}$ orbitals by local moments  coupled only through an interlayer $J_\bot$.
{Treating $J_\bot$ perturbatively}, we have employed a large-scale density-matrix renormalization group (DMRG) calculation and obtained a phase diagram in the $t'$-$U$ plane for a finite $J_H$, as shown in Fig.~\ref{fig:model}(c). The $(\pi/2,\pi/2)$ SSO consistent with experiments sets in only when $t'<t$ and $U$ is sufficiently large. It admits a simple explanation:  each one-dimensional (1D) `zig-zag' chain 
with hopping $t$ can be {\it ferromagnetic} due to double exchange from $J_H$~\cite{PhysRev.82.403,PhysRev.100.675}, and the weaker hopping $t'$ induces an anti-ferromagnetic  interchain  coupling that eventually leads to the $(\pi/2,\pi/2)$ spin stripe pattern [see Fig.~\ref{fig:model}(d)].
In the symmetric case when $t'=t$,
our DMRG result suggests that increasing $J_\bot$ enhances an interlayer $s$-wave pairing tendency\footnote{\ymw{Note these two sets of calculations should be viewed as representing two experimentally motivated structural regimes, rather than a controlled continuous interpolation along a single pressure- or strain-driven path in the model parameter space.}}.
Our study establishes a microscopic origin of the diagonal SSO in La$_3$Ni$_2$O$_7$ and shows that, within the same effective framework, stronger interlayer coupling in the more symmetric regime enhances interlayer pairing tendencies, highlighting the unique role of Hund's coupling in this new high-$T_c$ material.

{\it Model and method.}--\ymw{Apart from different spatial extensions, the tetragonal elongation on Ni sites further differentiates the $d_{z^2}$ orbital filling from the $d_{x^2-y^2}$ orbital filling even in the presence of strong interactions\cite{PhysRevLett.131.126001,PhysRevLett.132.146002,PhysRevB.111.L241102}.
This is consistent with recent theory and spectroscopy studies: dynamical-mean-field-theory finds local-moment behavior in the $d_{z^2}$ sector~\cite{PhysRevB.109.L081105}, optical data imply that low-energy carriers are dominately from $d_{x^2-y^2}$ component~\cite{Liu2024}
and ARPES finds a strongly renormalized flat $d_{z^2}$ band below the Fermi level~\cite{Yang2024}. Independently, we have investigated the ground state of a Hund-Hubbard model with both orbitals and realistic parameters, and found $\braket{n_{x^2-y^2}}=1/2$ and $\braket{n_{z^2}}=1$, as detailed in the Supplementary Materials (SM)\cite{SM}. Thus, it suffices to approximate the $d_{z^2}$ orbital using local moments and depart from the following effective Hamiltonian}  
\begin{equation}
     \begin{aligned}
         H=&-\sum_{\langle ij\rangle,\ell,\sigma} \left(t_{ij} c_{i,\ell,\sigma}^\dagger c_{j,\ell,\sigma}+h.c. \right) +U\sum_{i\ell}n_{i\ell \uparrow}n_{i\ell\downarrow}\\
         &-J_H \sum_{i\ell}\bm{s}_{i,\ell}\cdot\bm{S}_{i,\ell}+J_\bot\sum_i \bm{S}_{i,1}\cdot\bm{S}_{i,2}.
     \end{aligned}\label{eq:model}
 \end{equation}  
Here $\bm{s}_{i,\ell}=\frac{1}{2}\sum_{\alpha,\beta}c^\dagger_{i,\ell,\alpha}\bm{\sigma}_{\alpha\beta}c_{i,\ell,\beta}$ is the spin operator for the itinerant $d_{x^2-y^2}$ electrons at site $i$ in layer $\ell$ with $\bm{\sigma}$ the vector of Pauli matrices, $\bm{S}_{i,\ell}$ is the spin-1/2 operator for the localized $d_{z^2}$ electrons on the same site. $U$ is the Hubbard interaction for the itinerant electrons, $J_H>0$ is the Hund's coupling and $J_\bot>0$ is the interlayer super-exchange coupling between the local spins of the $d_{z^2}$ orbitals.
Only the nearest neighbor hoppings for $d_{x^2-y^2}$ orbitals are retained, with $t_{ij}=t$ on the strong bond, and $t_{ij}=t'$ on the weak bond, as shown in Fig.~\ref{fig:model}(b). 
Here $t'\neq t$ respects the $Amam$ structure at ambient pressure, and $t'=t$ in the $I4/mmm$ structure at high pressure (or with compressive strain in the thin film sample).
Similar models with non-uniform hopping $t_{ij}$ have also been used to study the impact of pressure on superconductivity in Ref.\cite{Lu2025}.  \ymw{We caution that if the $d_{z^2}$ sector becomes sufficiently itinerant, Eq.\eqref{eq:model} is no longer valid and explicit interlayer hopping must be retained; that regime lies beyond the scope of the present work.}
 Below, we will assume that at ambient pressure (or in  unstrained thin film), $J_\bot$ is not strong enough to bind the pair of interlayer $d_{z^2}$ orbitals into a singlet, such that the interlayer coupling can be treated perturbatively. This regime allows for a focus on one layer first, as the other layer is obtained by flipping the spin patterns. 
By contrast, in the high pressure regime, $J_{\perp}$ cannot be treated perturbatively and the DMRG treatment must include both orbitals in both layers.

The ground state is investigated via GPU accelerated  large-scale DMRG calculations~\cite{White1992} with a system size of up to 400 lattice sites, a bond dimension up to $D=20000$ and truncation error of the order $\epsilon\sim 10^{-6}$.
We employ the subspace expansion technique \cite{dmrg_subspace2015} while retaining the two-site update scheme to avoid  local minima.

\begin{figure}
    \includegraphics[width=\linewidth]{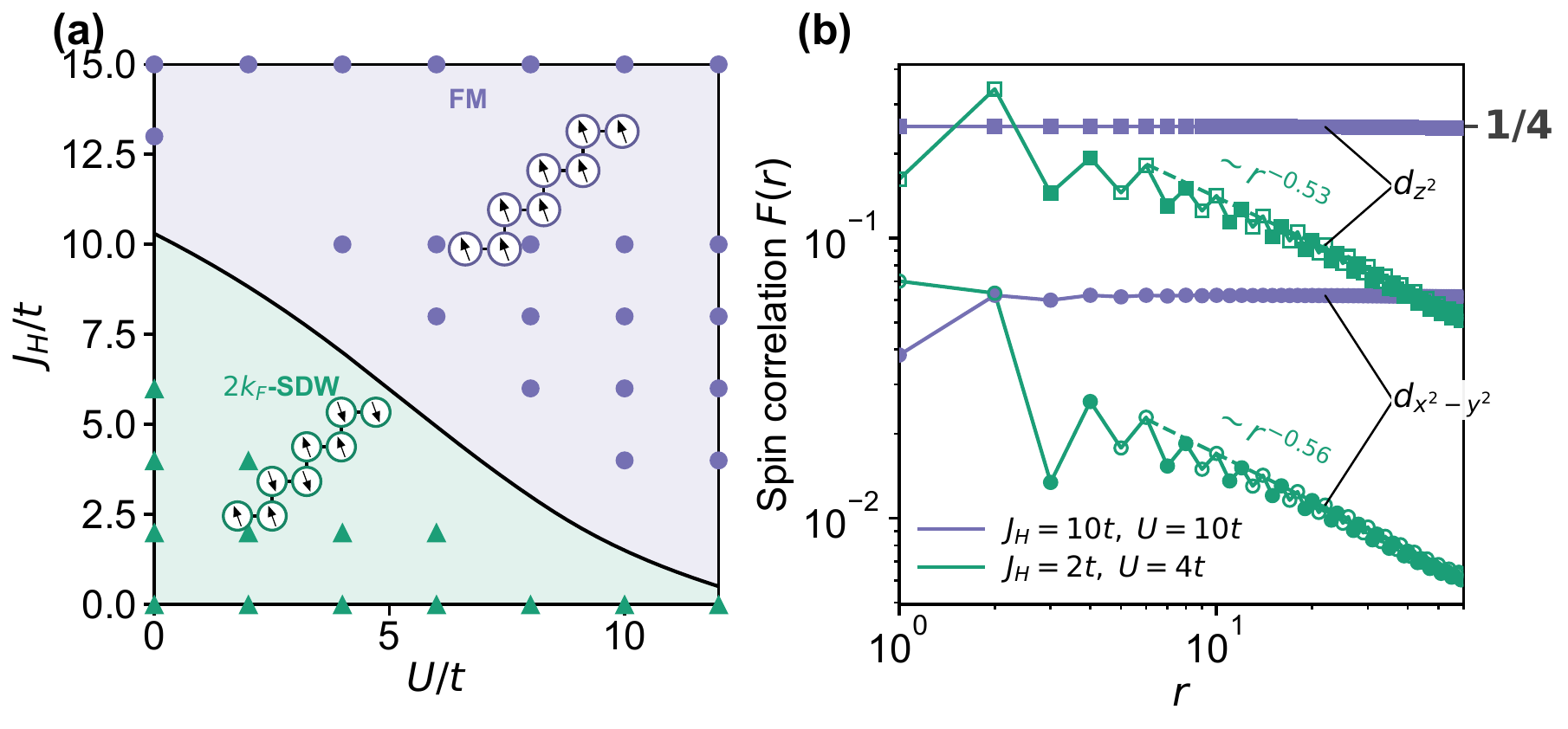}
    \caption{(a)  Phase diagram for the 1D model in Eq.~\eqref{eq:model2} at quarter filling ($\braket{n}=1/2$) determined by  the spin correlation. 
    \ymw{The two insets sketch the corresponding spin patterns along the chain: uniform polarization in the FM phase and a period-4 $\uparrow\uparrow\downarrow\downarrow$ pattern in the $2k_F$-SDW phase.} The FM and $2k_F$ orders can be adiabatically connected to the $(\pi/2,\pi/2)$ and $(0,\pi)$ SDW orders respectively in the 2D limit. (b) Representative spin correlation functions  for the FM and $2k_F$-SDW phases, \ymw{and for both the itinerant $d_{x^2-y^2}$ electrons (circles) and the localized $d_{z^2}$ moments (squares);}  filled (hollow) markers
    \ymw{indicate} positive (negative) values. \ymw{In the FM phase the $d_{z^2}$ correlation saturates at $\langle\bm{S}_i\!\cdot\!\bm{S}_j\rangle\!\to\!1/4$ (right axis), while in the $2k_F$-SDW phase both decay with essentially the same power law $r^{-\alpha}$ ($\alpha\!\approx\!0.55$), confirming that the spin orders for both orbitals are locked together by $J_H$.}
    }\label{fig:1Dphase}
\end{figure}

{\it Decoupling limit.}--Before discussing the spin stripes, it is helpful to first consider the artificial limit when $t'\to0$, i.e. the hopping on the weak bonds vanishes. If we focus on a single layer by ignoring $J_\perp$, the system reduces to a 1D `zig-zag' Kondo-Hubbard lattice along $45^\circ$-direction which is described by 
    \begin{equation}
     \begin{aligned}
         H=&- t\sum_{i,\sigma}  c_{i\sigma}^\dagger c_{i+1\sigma}+h.c.+U\sum_{i}n_{i \uparrow}n_{i\downarrow}-J_H \sum_{i}\bm{s}_{i}\cdot\bm{S}_{i}
     \end{aligned}\label{eq:model2}
 \end{equation} 
at $1/4$ filling. 
Much of the physics from this model is known when either $U$ or $J_H$ is absent. 

When $J_H$ is negligible Eq.\eqref{eq:model2} reduces to the 1D Hubbard model, for which the ground state cannot be ferromagnetic at any finite filling and any strength $U$, as is well known from the Lieb-Mattis theorem~\cite{PhysRev.125.164,lieb1962ordering}. In fact, the system tends to develop a $2k_F$ density wave order, which,  in the case of $1/4$ filling,  corresponds to period-4 commensurate density wave. Although both charge and spin density wave (CDW and SDW) exhibit the same power-law correlation, the latter is more favorable due to logarithmic corrections~\cite{JVoit_1988,PhysRevB.39.4620,Affleck_1989}. Thus, the ground state in this case is a period-4 $2k_F$-SDW.

In the other limit when $U$ is negligible, the model reduces to the  ferromagnetic Kondo model (also known as the double-exchange model) with $J_H$ playing the role of the Kondo coupling.
Interestingly, both the ferromagnetic ($J_H>0$, see Refs.~\cite{PhysRevB.58.6414,PhysRevB.72.075118,PhysRevB.61.9532,PhysRevB.64.012416}) and antiferromagnetic ($J_H<0$, see Refs.~\cite{PhysRevB.46.13838,PhysRevLett.78.2180,PhysRevB.58.2662,PhysRevLett.77.1342,PhysRevB.65.144419,PhysRevB.71.214415,RevModPhys.69.809,PhysRevB.86.165107,PhysRevB.47.2886}) 1D Kondo lattice model feature a {\it ferromagnetic} ground state at large $|J_H|$ with conduction electron density $n\lesssim0.8$. 
In particular when $J_H\to+\infty$, the system is equivalent to spinless itinerant electrons~\cite{PhysRevB.58.6414,PhysRevB.61.9532}, where all the spins are polarized.

When $J_H$ and $U$ are both present, there is a competition between the $2k_F$-SDW and ferromagnetic orders. To reveal the ground state of Eq.~\eqref{eq:model2} at arbitrary $(J_H,U)$, we calculated the spin-spin correlation function  $F(|\br_i-\br_j|) = \braket{\bm{s}_i\cdot\bm{s}_j}$ using DMRG, and identified different ordering tendencies and  the phase diagram in Fig.~\ref{fig:1Dphase}(a) with the insets indicating the spatial patterns in each phase. {
\ymw{Representative spin correlation functions for both the itinerant $d_{x^2-y^2}$ electrons and the localized $d_{z^2}$ moments are presented in Fig.~\ref{fig:1Dphase}(b).} \ymw{Due to the ferromagnetic nature of $J_H$, 
\ymw{the two orbitals exhibit the same correlation pattern in both phases.}} The correlation in the $2k_F$-SDW state exhibits a power-law decay with a sign-change pattern of $++--$, whereas that in the FM state shows long-range order without decay and takes positive values throughout.}
For a given  $U$, a large enough $J_H$ will always drives the system into a ferromagnetic ground state.
Since the superexchange process which requires virtual double occupancy and tends to destroy the FM ordering is suppressed at larger $U$, the critical $J_H$ marking the onset of FM ordering is smaller with a larger $U$. 
Indeed, our numerical results indicate that the critical value of $J_H$ decays with increasing $U$ [see Fig.~\ref{fig:1Dphase}(a)].  

\begin{figure}[bt]
    \includegraphics[width=0.9\linewidth]{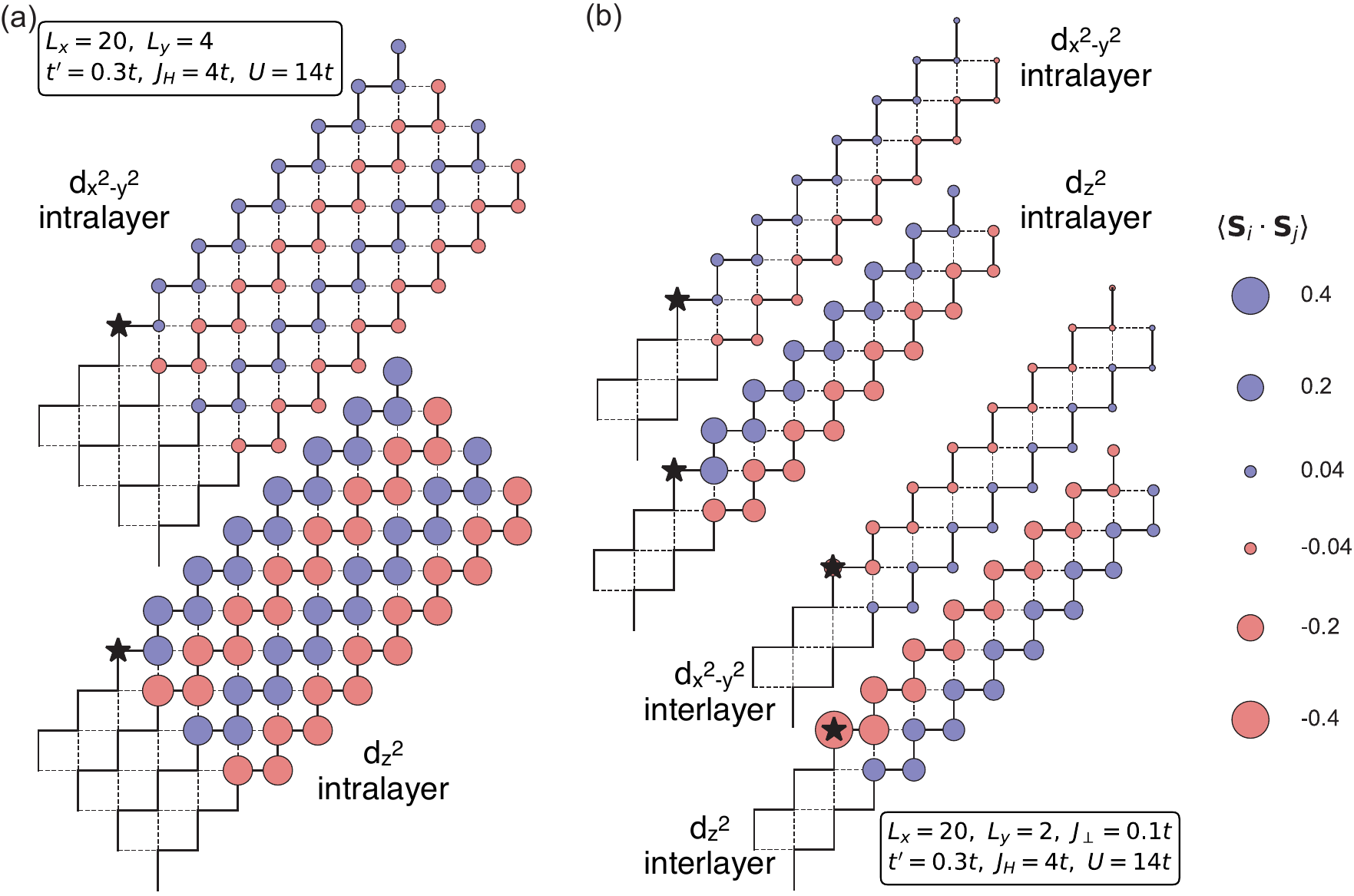}
    \caption{\ymw{(a) Spin correlation function $F(\br)$ from a four-chain, monolayer DMRG calculation, implying the $(\pi/2,\pi/2)$ order for both orbitals in the 2D limit. (b) $F(\br)$ from a two-chain, bilayer calculation. The interlayer pattern arises from the antiferromagnetic $J_\bot$. The insets show the used parameters,  the size and color of circles indicate the values of $F(\br)$ and the stars mark the reference position for $F(\br)$. }
    }\label{fig:twochain}
\end{figure}

{\it Coupled multi-chain system.}--Having identified the ground states for the decoupled chain system, we now consider the crossover to the 2D case by switching on a small $t'\neq0$ that couples the chains. 
Before presenting the DMRG results, some perturbative arguments in favor of the $(\pi/2,\pi/2)$ SSO are in order. Suppose the 1D system is in the ferromagnetic regime [see the purple region in Fig.~\ref{fig:1Dphase}(a)], a small $t'<U$ will induce anti-ferromagnetic coupling $J\sim t'^2/U$ between neighboring chains.
 \ymw{Note in the other limit $U\to0$ the interchain coupling could be ferromagnetic due to the short-distance RKKY interaction between local moments\cite{Chen2022PRB}, which, however, is not in the regime relevant to La$_3$Ni$_2$O$_7$.} 
 In this case, the more energetically favorable state is when the spin polarization in each chain alternates from one to another, eventually leading to the $(\pi/2,\pi/2)$ SSO in the 2D limit, as in Fig.~\ref{fig:model}(d). 

Indeed, this is confirmed by our DMRG results. In Fig.~\ref{fig:twochain}(a)  we show the calculated spin correlation function $F(\br)$ for both orbitals and for a set of representative parameters in a four-chain, single layer system \ymw{with open boundary condition (results with periodic boundary conditions exhibit the same behavior, as shown in the SM\cite{SM}).} Clearly, the pattern is consistent with the 2D $(\pi/2,\pi/2)$ SSO. Using the same strategy we have identified the regime in the $t'$-$U$ plane that stabilizes this order for $J_H=4t$, as shown in Fig.~\ref{fig:model}(c). Note that our theory of the $(\pi/2,\pi/2)$ SSO discussed here \ymw{does not require fine tuning of the electron filling} but crucially depends on the presence of a sizable $J_H$. 
We have confirmed numerically that by taking a smaller $J_H$, one needs a larger $U$ to stabilize the stripe order. 
This is consistent with our 1D-limit result that for a smaller $J_H$, a larger $U$ is required to stabilize the FM order.  
 In the limit of $J_H\to0$, the spin stripe of this kind would never arise, as we learn from Fig.~\ref{fig:1Dphase} (a) 
that each decoupled chain is always in a $2k_F$ order without $J_H$.

\ymw{Although the results discussed above are obtained from single layer calculations, adding a small $J_\bot$ and including both layers merely flips spin patterns between the two layers. We confirm this by calculating $F(\br)$ for a two-chain bilayer system, and the resutls are shown in Fig.~\ref{fig:twochain}(b). In particular, the {\it interlayer} $F(\br)$ for both orbitals clearly reflects the antiferromagnetic character of $J_\bot$.    }

\begin{figure}
    \includegraphics[width=\linewidth]{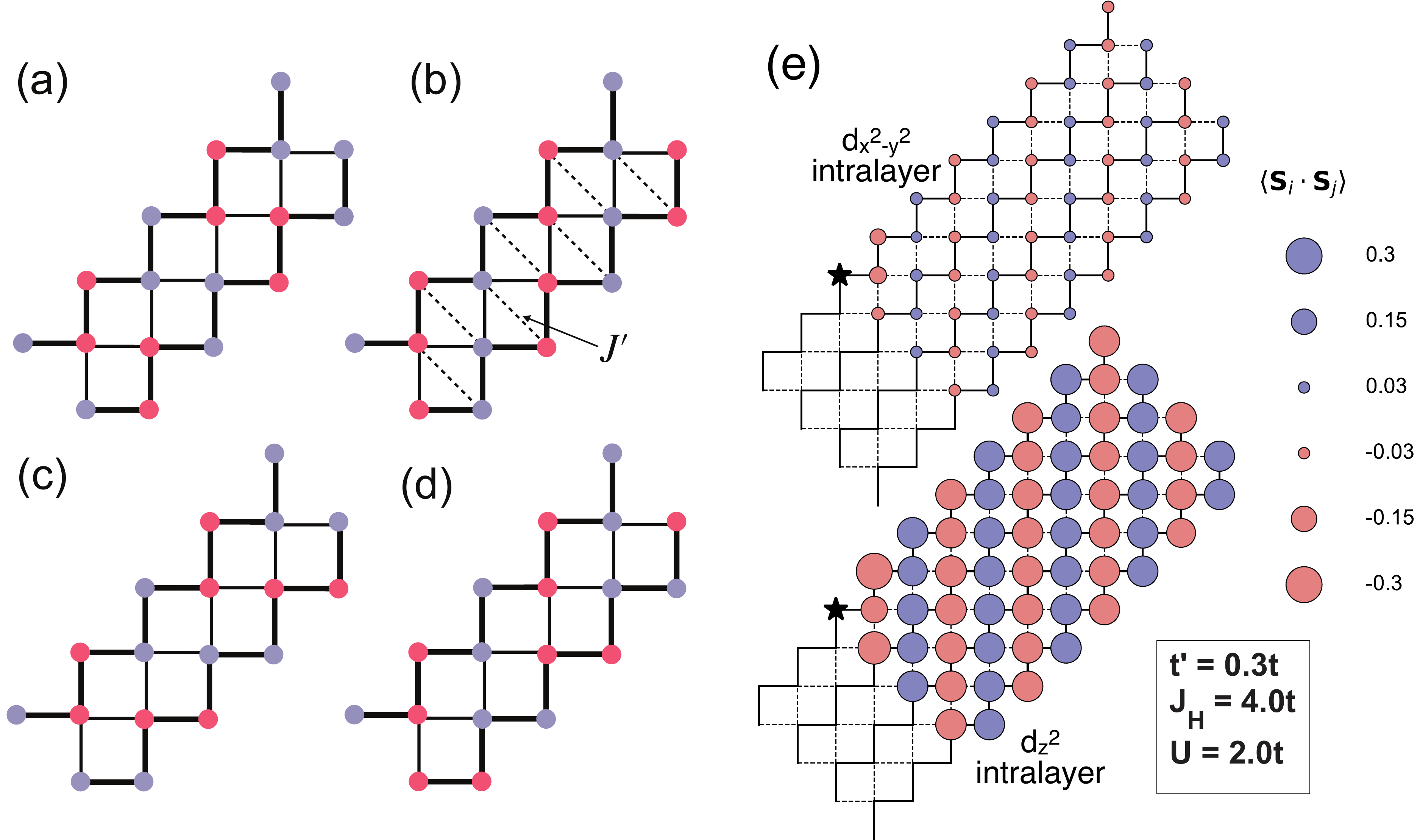}
    \caption{(a-d)Four different possibilities of spin pattern in the coupled two-chain system when each chain favors the period-4 SDW tendency. Among them, (a) corresponds to another type of $(\pi/2,\pi/2)$ SSO, while (b) corresponds to a $(\pi,0)$ SSO. Our DMRG results suggest the pattern in (b) is more stable, which can be explained through a fourth-order perturbation term $J'\sim t^2t'^2/U^3$ that couples diagonal sites in each square plaquette. (e) {Spin-spin correlation measured by DMRG with the shown parameters, confirming the $(\pi, 0)$ SSO.}}\label{fig:twochain2}
\end{figure}

There is another possible candidate for the $(\pi/2,\pi/2)$ SSO though, which comes from coupling period-4 SDW chains as depicted in  Fig.~\ref{fig:twochain2}(a) and thus may emerge in the absence of $J_H$. 
This type of stripe order will be sensitive to the filling fraction of the $d_{x^2-y^2}$ orbitals, \ymw{as the specific period-4 of the SDW is tied to the $1/4$ filling.}
But more importantly, this state is not energetically favored. Consider the two chain system as in Fig.~\ref{fig:twochain2}(a), sliding one chain relative to the other along the diagonal direction leads to four different patterns shown in Fig.~\ref{fig:twochain2}(a-d). If only the super-exchange inter-chain coupling (for $d_{x^2-y^2}$ orbitals) $J\sim t'^2/U$ is considered, all four patterns are equally frustrated. 
However, it is easy to see that a higher order exchange term $J'\sim (tt')^2/U^3$ coupling the chains along the plaquette diagonal direction stabilizes the pattern in Fig.~\ref{fig:twochain2}(b), corresponding to a $(\pi,0)$ order.
Indeed our DMRG calculation on both coupled-two-chain and coupled-four-chain ladders identifies a large parameter regime for this $(0,\pi)$ or $(\pi,0)$ SSO at either smaller $U$ or larger $t'$ [see the green region in Fig.\ref{fig:model}(c)]. The small $U$ result confirms our analysis above, as shown in Fig.~\ref{fig:twochain2}(e) where the calculated $F(\br)$ for both orbitals on a coupled-four-chain ladder is presented. 
While for the $t'\to t$ case the picture of coupled 1D chains  breaks down since $t'$ cannot be treated perturbatively.

\begin{figure}[htbp]
    \includegraphics[width=0.85\linewidth]{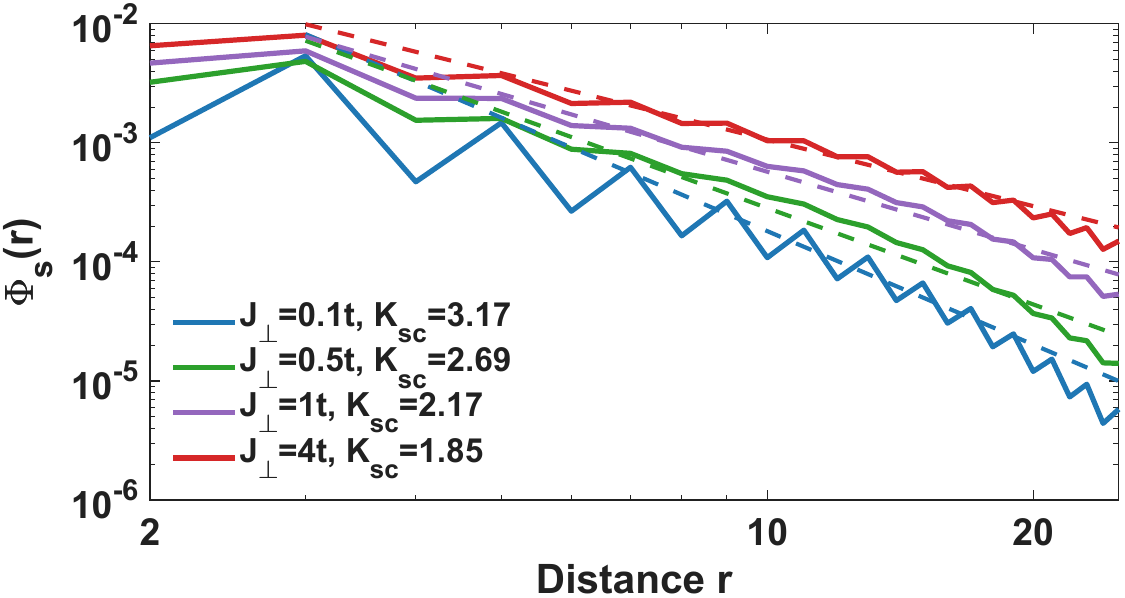}
    \caption{Interlayer singlet pair correlation function $\Phi_s(\br)$ calculated from Eq.~\eqref{eq:model} on a regular two layer, two leg and two orbital system for $J_H=4t,  U=18t$.
    }\label{fig:interpairing}
\end{figure}

\ymw{{\it Pairing tendencies at $t'=t$.}--}In the compressively strained thin film sample, or approximately in the bulk sample at high pressure where superconductivity is observed, the system retains tetragonal symmetry, implying $t'=t$ in our model. To inspect the ground state in this case, we perform DMRG calculations on a regular square lattice with two layers, two orbitals and two legs, for the model in Eq.~\eqref{eq:model} with $t_{ij}=t$ \ymw{and $J_\perp$ explicitly included, as motivated by structural and theoretical studies indicating that high pressure enhances the interlayer $d_{z^2}$ overlap and hence the effective $J_\perp$~\cite{Wang2024,GeislerStruct2024,JiangPressure2025,Lu2025}.} 
\ymw{To diagnose the leading pairing tendency, we calculate the interlayer singlet pair correlation}
$\Phi_s(\br)=\braket{\Delta^\dagger(\bm{0}) \Delta(\br)}$, where $\Delta(\br_i)=(c_{i,1, \uparrow} c_{i,2,\downarrow } - c_{i,1,\downarrow} c_{i,2,\uparrow})/{\sqrt{2}}$ for   interlayer singlet $s$-wave pairing~\footnote{Note that in the presence of a finite coupling between $d_{x^2-y^2}$ and $d_{z^2}$ orbitals (which has been neglected in our model), the $d_{x^2-y^2}$ Fermi surface will split into bonding and anti-bonding parts, on which the gap function changes sign, thus corresponding to $s_\pm$-wave pairing}. The intralayer singlet pairing and interlayer triplet pairing are also investigated. However, we found that the intralayer pairing decays exponentially fast, while the interlayer triplet pairing is two-orders of magnitude smaller than the interlayer singlet pairing.  \ymw{Therefore, we focus on $\Phi_s(\br)$ defined above.} In Fig.~\ref{fig:interpairing} we present the calculated  $\Phi_s(\br)$ for different values of $J_\perp$ with other parameters fixed.  
 {It is clear that $\Phi_s(\br)$ can exhibit power-law decay $r^{-K_{sc}}$ when the interlayer coupling $J_\bot$ becomes large.
Specifically,  the exponent $K_{sc} < 2$ when $J_\bot = 4t$ for $J_H=4t$, indicating the divergence of SC susceptibility.} 
 \ymw{Within this framework, increasing $J_\bot$ by high pressure provides a natural route toward stronger interlayer pairing tendencies in this bilayer system,} consistent with the previous results using effective one orbital model or type-II $t-J$ model~\cite{PhysRevB.110.104517,PhysRevLett.132.036502} where similar $s$-wave pairing is also obtained.

{\it Discussion.}--Several candidate stripe orders have been proposed to explain the experimental data. In addition to the $(\pi/2,\pi/2)$ SSO discussed in our work [see Fig.~\ref{fig:model}(d)], another alternating spin-charge order also seems consistent with experiment in which Ni$^{2+}$ and Ni$^{3+}$ cations form alternating stripe structure and spin stripes only reside on the Ni$^{2+}$-stripe (see \cite{Chen2024} for details, where our SSO is referred to as double spin stripe order).
On the other hand, NMR measurement on the La-sites found no evidence of charge orders so far~\cite{ZHAO20251239}, probably due to the limit of sensitivity of charge orders on the La-atoms. 
From a symmetry perspective, such a charge order with a momentum $\bq_c=(\pi,\pi)$ is certainly  allowed, and can actually be induced from the $(\pi/2,\pi/2)$ SSO.
This can be seen from the following Ginzburg-Landau free energy
\begin{equation}
    \mathcal{F}=\alpha |\rho_{{\bq_c}}|^2+\lambda [\rho_{{\bq_c}}(S^i_{-{\bQ}})^2+c.c.],
\end{equation}
where $\alpha>0$ meaning no spontaneous transition to the CDW order, and to preserve translational symmetry we must have $\bq_c=2\bQ$. It is easy to see that with a finite $S^i_{\pm\bQ}\neq0$, $\mathcal{F}$ is  minimized at a nonzero charge order $\rho_{\bq_c}=-\frac{\lambda}{\alpha}(S^i_{\bQ})^2$. Therefore, although we have found a double spin stripe order as the ground state, charge order can still set in as it obeys all the symmetries of the system and can be induced from spin stripes.

{\it Acknowledgements}.-- We are grateful to Steven Kivelson, Zhengcheng Gu and Yahui Zhang for insightful discussions.  H.O. is supported by a startup
fund from Johns Hopkins University.  T.H., Y.Y., H.H., H.-C.J., Y.W.
and S.R. are supported by the US Department
of Energy, Office of Basic Energy Sciences, Division of
Materials Sciences and Engineering, under Contract No.
DE-AC02-76SF00515. Y.W. acknowledges support from
the Gordon and Betty Moore Foundation’s EPiQS Initiative through GBMF8686.
T.H. was supported by the Deutsche Forschungsgemeinschaft (DFG, German
Research Foundation) under Project No. 537357978.

\bibliography{nickelate}

\end{document}


\title{Supplementary Materials:\\
Origin of Spin Stripes in Bilayer Nickelate La$_3$Ni$_2$O$_7$}
    \author{Hao-Xin Wang}
    \affiliation{Department of Physics, The Chinese University of Hong Kong, Sha Tin, New Territories, Hong Kong, China}
    \author{Hanbit Oh}
    \affiliation{William H. Miller III Department of Physics and Astronomy, Johns Hopkins University, Baltimore, Maryland, 21218, USA}
    \author{Tobias Helbig}
    \affiliation{Leinweber Institute for Theoretical Physics, Stanford University, Stanford, CA 94305, USA}
    \author{Bai Yang Wang}
    \affiliation{Stanford Institute for Materials and Energy Sciences, SLAC National Accelerator Laboratory, Menlo Park, CA 94025, USA}
    \affiliation{Departments of Applied Physics, Stanford University, Stanford, CA 94305, USA}
    \author{Jiarui Li}
    \affiliation{Stanford Institute for Materials and Energy Sciences, SLAC National Accelerator Laboratory, Menlo Park, CA 94025, USA}
    \affiliation{Departments of Applied Physics, Stanford University, Stanford, CA 94305, USA}
    \author{Yijun Yu}
    \affiliation{Stanford Institute for Materials and Energy Sciences, SLAC National Accelerator Laboratory, Menlo Park, CA 94025, USA}
    \affiliation{Departments of Applied Physics, Stanford University, Stanford, CA 94305, USA}
    \author{Harold Y. Hwang}
    \affiliation{Stanford Institute for Materials and Energy Sciences, SLAC National Accelerator Laboratory, Menlo Park, CA 94025, USA}
    \affiliation{Departments of Applied Physics, Stanford University, Stanford, CA 94305, USA}
    \author{Hong-Chen Jiang}
    \affiliation{Stanford Institute for Materials and Energy Sciences, SLAC National Accelerator Laboratory, Menlo Park, CA 94025, USA}
    \author{Yi-Ming Wu}
    \affiliation{Leinweber Institute for Theoretical Physics, Stanford University, Stanford, CA 94305, USA}
    \affiliation{Stanford Institute for Materials and Energy Sciences, SLAC National Accelerator Laboratory, Menlo Park, CA 94025, USA}
    \author{S. Raghu}
    \affiliation{Leinweber Institute for Theoretical Physics, Stanford University, Stanford, CA 94305, USA}
    \affiliation{Stanford Institute for Materials and Energy Sciences, SLAC National Accelerator Laboratory, Menlo Park, CA 94025, USA}

\begin{abstract}
    In this Supplementary Material, we show that i) explicit two-orbital DMRG calculations in the electron coordinate confirms the local moment approximation for the $d_{z^2}$ orbitals and ii) our DMRG results are robust against using different boundary conditions. 
\end{abstract}

\maketitle

\section{Two-orbital DMRG in the electron coordinate}

To justify the local moment approximation for the $d_{z^2}$ orbitals that has been used throughout the main text, here we provide additional DMRG  results obtained using electron coordinate for both orbitals. Since the essential features of the material are multi-orbit and strong interaction, it suffices to consider the following simplified bilayer two-orbital Hund-Hubbard model,
\begin{equation}
     \begin{aligned}
         H=&-\sum_{\langle ij\rangle,\ell,\sigma} t_{ij}
         \left(c_{i\ell \sigma}^\dagger c_{j\ell \sigma}+h.c. \right)
         -t_\bot\sum_{i,\sigma}\left(f_{i1\sigma}^\dagger f_{i2\sigma}+h.c.\right) \\
         &+U\sum_{i,\ell,m=c,f}n_{i\ell m \uparrow}n_{i\ell m\downarrow}
         -J_H \sum_{i\ell}\bm{S}_{i\ell c}\cdot\bm{S}_{i\ell f},
     \end{aligned}\label{eq:all_dof_model}
 \end{equation}
in which both Ni-derived $e_g$ orbitals are retained explicitly. Here $c^\dagger_{i\ell \sigma}$ creates an $d_{x^2-y^2}$ electron on site $i$, layer $\ell=1,2$ and spin $\sigma$, and $f^\dagger_{i\ell \sigma}$ creates an $d_{z^2}$ electron on site $i$, layer $\ell=1,2$ and spin $\sigma$. The density and spin operators are
\begin{equation}
    \begin{aligned}
        &n_{i\ell c\sigma}=c^\dagger_{i\ell \sigma}c_{i\ell \sigma}, ~~ n_{i\ell f\sigma}=f^\dagger_{i\ell \sigma}f_{i\ell \sigma}\\
        &\bm{S}_{i\ell c}=\frac{1}{2}\sum_{\alpha,\beta}c^\dagger_{i\ell \alpha}\bm{\sigma}_{\alpha\beta}c_{i\ell \beta}, ~~\bm{S}_{i\ell f}=\frac{1}{2}\sum_{\alpha,\beta}f^\dagger_{i\ell \alpha}\bm{\sigma}_{\alpha\beta}f_{i\ell \beta}
    \end{aligned}
\end{equation}
The first term in Eq.\eqref{eq:all_dof_model} retains the in-plane hopping of the $d_{x^2-y^2}$ orbital with bond-dependent amplitude $t_{ij}$, while the second term keeps the explicit interlayer $d_{z^2}$-orbital hopping $t_\bot$.
Without loss of generality, we focus on the high-pressure phase where $t_{ij}\equiv t$ is homogeneous, and adopt the parameters from the DFT/Wannier study by Zhang {\it et al.}[55], which gives $t_{ij}=0.56\,{\rm eV}$ and $t_\bot=0.72\,{\rm eV}$,
i.e. $t_\bot/t=1.286$. For the interaction part, we consider the following two representative parameter sets in the correlated high-pressure window [37,55,59],
\begin{equation}
    \begin{aligned}
        &1.~~ U=4\,{\rm eV}, ~J_H=0.8\,{\rm eV},\\
&2. ~~ U=3\,{\rm eV}, ~J_H=0.8\,{\rm eV}.
    \end{aligned}\nonumber
\end{equation}
More precisely, this explicit check retains the intraorbital Hubbard repulsion and onsite Hund exchange needed for the orbital-selective question at hand, rather than the full rotationally invariant Kanamori interaction. In all calculations, we fix the total filling to $1.5$ electrons per Ni site and use an $L_x=6$, $L_y=2$ cluster with open boundary conditions and bond dimension $D=18000$.

The resulting site-averaged local observables for both orbitals are collected in Table~\ref{tab:all_dof_summary}.
Here $P_{{\rm single},m}=\langle n_{m\uparrow}(1-n_{m\downarrow})+n_{m\downarrow}(1-n_{m\uparrow})\rangle=\langle n_m\rangle-2\langle n_{m\uparrow}n_{m\downarrow}\rangle$ denotes the single-occupancy weight of orbital $m$, i.e. the probability that the orbital is occupied by exactly one electron. Physically, a larger $P_{{\rm single}}$ signals a stronger local-moment character. For the main high-pressure anchor ($U=4\,{\rm eV}$, $J_H=0.8\,{\rm eV}$), the $d_{z^2}$ sector is pinned to half filling, with $\langle n_{z^2}\rangle=1.0000$, a large single-occupancy weight $P_{{\rm single},z^2}=0.9199$, and a comparatively small charge fluctuation $\delta n_{z^2}^2=0.0801$. At the lower-$U$ anchor ($U=3\,{\rm eV}$, $J_H=0.8\,{\rm eV}$), the $d_{z^2}$ sector remains essentially half filled, with $\langle n_{z^2}\rangle=1.0000$, $P_{{\rm single},z^2}=0.8799$, and $\delta n_{z^2}^2=0.1201$. By contrast, the itinerant $d_{x^2-y^2}$ sector stays close to quarter filling and is substantially more charge fluctuating in both cases, with $\delta n_{x^2-y^2}^2=0.2666$ and $0.2743$, respectively. Therefore, these calculations show that explicit interlayer $d_{z^2}$ hopping still leaves the $d_{z^2}$ electrons in the strongly correlated, near-local-moment regime suitable for the reduced $J_\bot$ description.

\begin{figure}
    \centering
    \includegraphics[width=\linewidth]{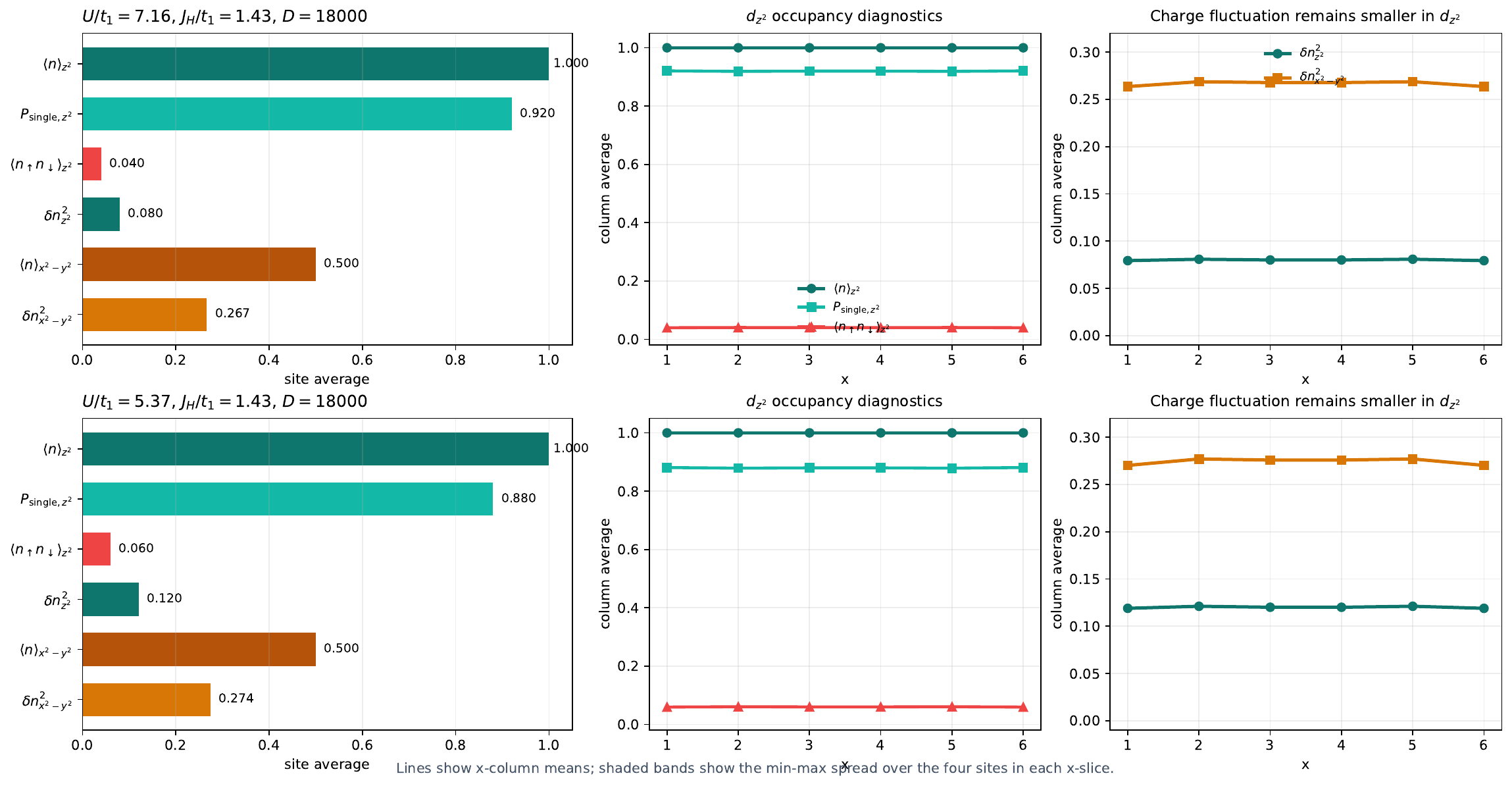}
\caption{Comparison of the two highest-bond-dimension explicit two-orbital datasets used to assess the localized $d_{z^2}$ regime with explicit interlayer hopping. Top row: main high-pressure anchor, $U/t_1=7.16$ and $J_H/t_1=1.43$ ($U=4\,{\rm eV}$, $J_H=0.8\,{\rm eV}$). Bottom row: lower-$U$ high-pressure anchor, $U/t_1=5.37$ and $J_H/t_1=1.43$ ($U=3\,{\rm eV}$, $J_H=0.8\,{\rm eV}$). Left column: site-averaged one-site diagnostics. Middle column: $x$-resolved $d_{z^2}$ occupancy diagnostics. Right column: $x$-resolved charge fluctuations. In both realistic $J_H=0.8\,{\rm eV}$ anchors the $d_{z^2}$ sector stays essentially half filled and close to the local-moment limit.}
    \label{fig:all_dof_compare}
\end{figure}

Figure~\ref{fig:all_dof_compare} visualizes the full real-space-resolved ($x$-direction) one-site diagnostics of these two calculations. The lines show the layer- and $L_y$-averaged values at each $x$ slice.
We see that the occupation number for the $d_{z^2}$ orbitals stays uniformly close to 1 in both rows of Fig.~\ref{fig:all_dof_compare}. The lower-$U$ case shows somewhat larger charge fluctuations than the $U=4\,{\rm eV}$ anchor, but in both cases the $d_{z^2}$ sector remains the less charge-fluctuating one. This provides direct justification for the low-energy local moment approximation adopted in the main text, since the $d_{z^2}$ orbitals stay in the correlated, near-half-filled regime rather than an itinerant limit.

\begin{table*}[t]
    \centering
    \resizebox{\linewidth}{!}{
    \begin{tabular}{c c c c c c c c c c}
    \hline\hline
    $U/t_1$ & $J_H/t_1$ & $\langle n \rangle_{z^2}$ & $\langle n_\uparrow n_\downarrow \rangle_{z^2}$ & $P_{{\rm single},z^2}$ & $\delta n_{z^2}^2$ & $\langle n \rangle_{x^2-y^2}$ & $\langle n_\uparrow n_\downarrow \rangle_{x^2-y^2}$ & $P_{{\rm single},x^2-y^2}$ & $\delta n_{x^2-y^2}^2$ \\
    \hline
    7.1556 & 1.4311 & 1.0000 & 0.0401 & 0.9199 & 0.0801 & 0.5000 & 0.0083 & 0.4834 & 0.2666 \\
    5.3667 & 1.4311 & 1.0000 & 0.0600 & 0.8799 & 0.1201 & 0.5000 & 0.0122 & 0.4757 & 0.2743 \\
    \hline\hline
    \end{tabular}}
    \caption{Site-averaged one-site observables for the two highest-bond-dimension ($D=18000$) explicit two-orbital DMRG datasets discussed in this section. Both runs use the same $L_x=6$, $L_y=2$ OBC cluster with explicit interlayer $d_{z^2}$ hopping. Throughout this table, energies are quoted in units of the DFT/Wannier in-plane hopping $t_1=0.56\,{\rm eV}$; the corresponding interlayer hopping is $t_\bot=0.72\,{\rm eV}$, i.e. $t_\bot/t_1=1.286$. The two rows correspond to the realistic high-pressure interaction window $J_H=0.8\,{\rm eV}$ with $U=4\,{\rm eV}$ and $3\,{\rm eV}$, respectively.}
    \label{tab:all_dof_summary}
\end{table*}

\section{Robustness of the stripe order: cylindrical boundary conditions}

In the main text, all multi-chain DMRG calculations were performed with open boundary conditions (OBC). Here we verify that both the $(\pi/2,\pi/2)$ stripe and the $(\pi,0)$ columnar magnetic orders are robust under cylindrical boundary conditions (periodic along the transverse direction, open along the longitudinal direction) on the four-chain ladder.

For the four-chain ladder on our 45$^\circ$-oriented zigzag lattice, cylindrical boundary conditions provide a nontrivial check by eliminating edge effects along the direction perpendicular to the chain. In particular, the periodic wrapping introduces additional diagonal bonds that are absent under OBC, so confirming the same ordering pattern on the cylinder rules out boundary artifacts.

\begin{figure}
    \centering
    \includegraphics[width=\linewidth]{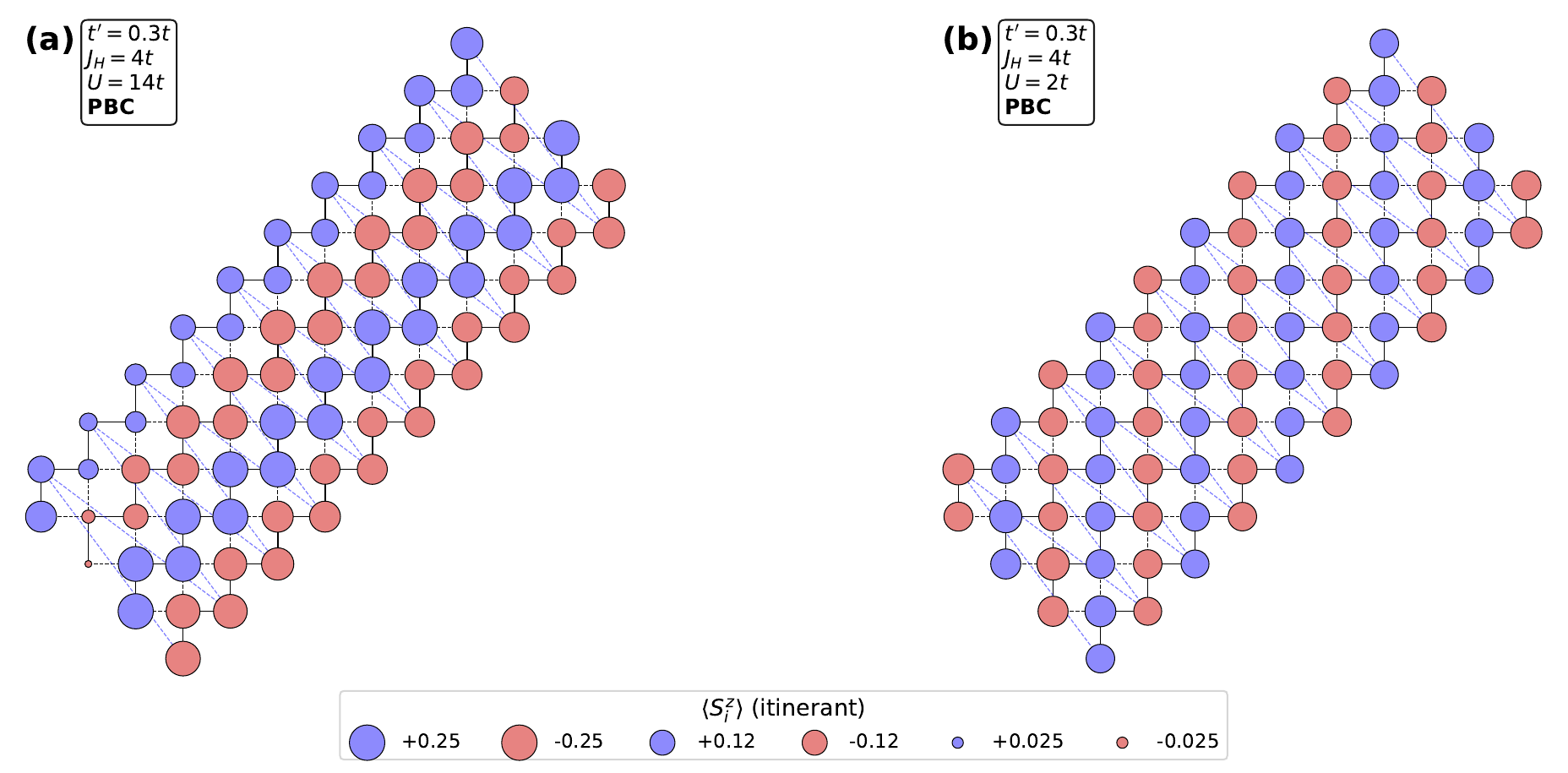}
    \caption{Local magnetization $\langle S^z_i\rangle$ of the itinerant $d_{x^2-y^2}$ electrons on the four-chain Kondo ladder ($L_y=4$, $L_x=20$) with cylindrical boundary conditions. (a)~Stripe regime: $t'=0.3t$, $J_H=4t$, $U=14t$, $D=12000$. The diagonal $(\pi/2,\pi/2)$ pattern is clearly visible. (b)~Columnar regime: $t'=0.3t$, $J_H=4t$, $U=2t$, $D=14000$. The $(\pi,0)$ pattern is robust on the cylinder.}
    \label{fig:pbc_halfpi}
\end{figure}

Figure~\ref{fig:pbc_halfpi} shows the local magnetization $\langle S^z_i\rangle$ of the itinerant $d_{x^2-y^2}$ electrons on the four-chain cylinder for both parameter regimes. In the stripe regime [panel~(a), $U=14t$, bond dimension $D=12000$], the $(\pi/2,\pi/2)$ diagonal stripe pattern is clearly reproduced: the magnetization exhibits the characteristic sign pattern with ferromagnetic alignment along the zigzag chains and antiferromagnetic alternation between neighboring chains, consistent with the OBC results in Fig.~3 of the main text. In the columnar regime [panel~(b), $U=2t$, $D=14000$], the $(\pi,0)$ pattern is likewise robust on the cylinder, consistent with Fig.~4(e) of the main text. We note that the bond dimension used in these cylindrical calculations ($D=12000$ for the stripe regime) is not yet fully converged---cylinders require substantially larger $D$ than open ladders for the same system size due to the increased entanglement---but the magnetic ordering pattern is already evident and unambiguous.

